\documentclass[usletter, 10pt, conference]{IEEEtran}

\usepackage{cite}

\usepackage[utf8]{inputenc}
\usepackage{times}

\usepackage[T1]{fontenc}

\usepackage[pdftex, bookmarks=false, draft]{hyperref}
\usepackage{graphicx}
\DeclareGraphicsExtensions{.pdf,.jpeg,.png}
\usepackage{verbatim}  
\usepackage{listings}  

\usepackage{amsmath}

\usepackage{algorithmic}

\usepackage[caption=false,font=footnotesize]{subfig}

\usepackage{url}

\usepackage{booktabs}
\usepackage{tabularx}
\usepackage[tracking,kerning,spacing]{microtype}

\usepackage[english]{babel}

\begin{document}
%
\title{	Demonstration of a Context-Switch Method\\ 
	for Heterogeneous Reconfigurable Systems
      }

\author{
	\IEEEauthorblockN{Arief Wicaksana, 
			  Alban Bourge, 
			  Olivier Muller, 
			  Frédéric Rousseau}
	\IEEEauthorblockA{
                TIMA Laboratory - Grenoble-INP/UGA/CNRS\\
                46 avenue Félix Viallet, 38000 Grenoble, FRANCE \\
                \{arief.wicaksana,alban.bourge,olivier.muller,frederic.rousseau\}@imag.fr
        	}
	}

\maketitle



\section{Introduction}
\label{sec:introduction}
Nowadays, FPGAs are integrated in high-performance computing
systems, servers, or even used as accelerators in System-on-Chip (SoC) platforms. 
Since the execution is performed in hardware, FPGA gives much higher performance
and lower energy consumption compared to most microprocessor-based systems.
However, the room to improve FPGA performance still exists, e.g. when it is used 
by multiple users.
In multi-user approaches, FPGA resources are shared between several users.
Therefore, one must be able to interrupt a running circuit
at any given time and continue the task at will.
An image of the state of the running circuit (context) is
saved during interruption and restored when the execution is continued.
The ability to extract and restore the context is known as context-switch.

In the previous work \cite{bourge_automatic_2015}, an automatic checkpoint selection
method is proposed for circuit generation targeting reconfigurable systems.
The method relies on static analysis of the finite state machine of a circuit
to select the checkpoint states. States with minimum overhead
will be selected as checkpoints, which allow optimal context save
and restore. The maximum time to reach a checkpoint
will be defined by the user and considered as the context-switch latency.
The method is implemented in C code and integrated as plugin in a free and open-source
High-Level Synthesis tool AUGH \cite{prost-boucle_fast_2014}.

\section{Demonstration}
In this demonstration, we present the context-switch method \cite{bourge_automatic_2015}
implemented in heterogeneous reconfigurable systems using a network-connected framework.
SoC-FPGA platforms with a CPU tightly-coupled with an FPGA are being used in the framework.
The demonstration framework consists of two FPGA-SoC platforms, a server with 
tool-chain installed and a network storage disk.
We use two different platforms in the framework, a ZC706 Evaluation Board from Xilinx and
an Arria V SoC Development Kit from Altera, to show the genericity of our method
even for FPGAs from different vendors. The server provides a tool-chain suite of
corresponding platforms for configuration purpose and the network storage is
implemented with NFS protocol.

A graphical and a non-graphical applications will be used to perform context-switch on the boards. These applications will consume test vectors as input and perform the computations.
For the graphical application, an output video will be prepared for each board so attendees can follow from the screen. 
For the non-graphical application, the flow of the execution will be observed from user's PC connected to the server.

\begin{figure}[tb]
        \centering
        \includegraphics[width=0.42\textwidth]{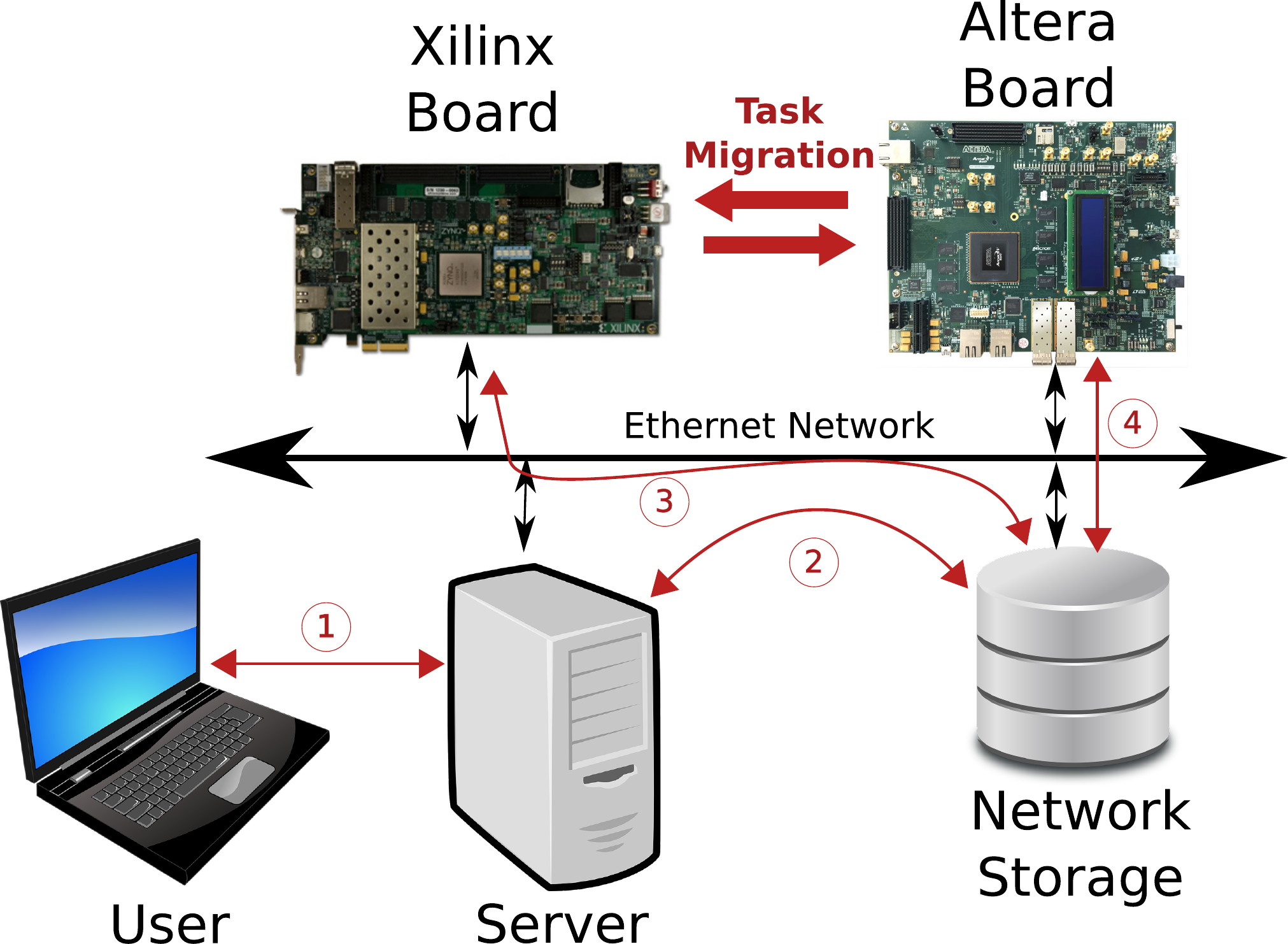}
	\raggedleft
        \caption{Demonstration Framework and its Typical Execution Flow}
        \label{fig:system}
         \vspace{-3mm}
\end{figure}

Figure \ref{fig:system} shows the general overview of our demonstration framework and
the execution flow.
A typical execution flow is given below. Other variations of boards
and steps are possible.
\begin{enumerate}
\item A user connects to \emph{Server} either via local connection or framework's network, put its
application and launch the execution.

\item The configuration of the circuit is generated in \emph{Server} and saved in \emph{Network
Storage} with its respective input test vectors.

\item The CPU on \emph{Xilinx Board} which detects the configuration file in 
\emph{Network Storage} and the
test vectors will program the FPGA and launch the execution. When there is an interruption,
it will retrieve the context from FPGA and save it in \emph{Network Storage}.

\item The CPU on \emph{Altera Board} which detects the bitstream and the context in \emph{Network Storage} will
program the FPGA and continue the execution. When there is no other interruption, it will finish the
execution and put the results in \emph{Network Storage}.
\end{enumerate}

\bibliographystyle{IEEEtran}
\bibliography{references}

\end{document}